 \documentstyle[epsf,amsfonts,amssymb]{article} 

\def\non{\nonumber}

\def\prtl{\partial} 
\def\para{{/\!\!/}} 
\def\half{\frac{1}{2}}

\def\cA{{\cal A}}  
\def\cB{{\cal B}}

\def\cK{{\cal K}}  
\def\cL{{\cal L}}  
\def\cM{{\cal M}}  
\def\cN{{\cal N}}  
\def\cO{{\cal O}} 
\def\cQ{{\cal Q}}  
\def\cR{{\cal R}} 
\def\cS{{\cal S}} 
\def\cT{{\cal T}} 
\def\cW{{\cal W}}

\def\Frac(#1/#2){\left(\frac{#1}{#2}\right)} 
\def\z2{{${\Bbb Z}_2$}} 
\def\al{{\alpha}}
\def\alb{{\alpha_B}}
\def\ds{{\mbox{dS}}}
\def\ads{{\mbox{adS}}} 
\def\min{{\mbox{{${\Bbb E}$}}}} 
\def\dsds(#1/#2){{dS$_{#1}$-brane $\subset$}{\,dS$_{#2}$-bulk}} 
\def\dsads(#1/#2){{dS$_{#1}$-brane $\subset$}{\,adS$_{#2}$-bulk}} 
\def\dsmin(#1/#2){{dS$_{#1}$-brane $\subset$}{\,${\Bbb E}_{#2}$-bulk}} 
\def\minads(#1/#2){{${\Bbb E}_{#1}$-brane $\subset$}{\,adS$_{#2}$-bulk}} 
\def\adsads(#1/#2){{adS$_{#1}$-brane $\subset$}{\,adS$_{#2}$-bulk}}

\newcommand{\vs}{\vspace*{5mm}}  
\newcommand{\svs}{\vspace*{3mm}}

\begin{document}
\def\abstract#1
{\begin{center}{\large \bf Abstract}\par #1 \end{center}}
\def\title#1{\begin{center}{{\Large \bf #1}}\end{center}}
\def\author#1{\begin{center}{{\large  #1}}\end{center}}
\def\address#1{\begin{center}{\large \it #1}\end{center}} 

\begin{flushright}    
YITP-02-48 \\        
\end{flushright}      
\vskip 0.3cm      

\today

\vskip 1.5cm      

\title{ Can a brane fluctuate freely? }
 
\vskip 0.3cm      
\author{Akihiro Ishibashi and Takahiro Tanaka} 

\vskip 0.3cm      

\address{
Yukawa Institute for Theoretical Physics, Kyoto University 
Kyoto 606-8502, Japan
}

\vs \svs 
\begin{center} {\large Abstract} \end{center}

\begin{center} 
\begin{minipage}{14cm}
 {
  No, it cannot in the following sense if a self-gravitating vacuum brane 
  is concerned. Once we write down the full set of linear perturbation 
  equations of the system containing a self-gravitating brane,   
  we will see that such a brane does not have its own dynamical degrees 
  of freedom independent of those of gravitational waves which propagate 
  in the surrounding spacetime. This statement seems to contradict with 
  our intuition that a brane fluctuates freely on a given background spacetime 
  in the lowest order approximation. 
  Based on this intuition, we usually think that the dynamics of a brane 
  can be approximately described by the equations derived from 
  the Nambu-Goto action. 
  In this paper we fill the gap residing between these two descriptions, 
  showing that the dynamics of a self-gravitating brane is in fact similar 
  to that described by a non-gravitating brane on a fixed background spacetime 
  when the weak backreaction condition we propose in this paper 
  is satisfied. 
          } 
\end{minipage}
\end{center}

\newpage 

\section{Introduction} 
\label{sect:introduction} 

Branes, general relativistic objects extended in spacetime, 
have been discussed in various contexts. 
In String or M-theory~\cite{text:Polchinski}, branes are thought to be 
a fundamental constituent element. In particular, D-branes on which 
the endpoints of open strings are restricted to lie have turned out 
to play an important role. 
In cosmology, topological defects~\cite{text:Vilenkin-Shellard} 
such as domain walls and cosmic strings may have influenced 
the history of our universe significantly. 
Recently, so called braneworld scenario has attracted much attention, 
giving an intriguing perspective that our universe is embedded as a
brane in a higher dimensional spacetime. 
In particular, in a model proposed by Randall and Sundrum~\cite{RS1999B}, 
the self-gravity of a self-gravitating tension brane plays an important role 
to reproduce the conventional law of gravity approximately. 

The dynamics of branes in a curved spacetime has so far been investigated 
extensively. 
One handy approximation to describe the dynamics of a brane 
is just to ignore its self-gravity, and to treat it as 
a test membrane on a given background spacetime. 
As far as a vacuum brane is concerned, 
its equation of motion follows from the Nambu-Goto action.   
The brane is supposed to oscillate freely following this equation 
in the lowest order approximation in which the self-gravity is neglected.  
Then, turning on the effect of self-gravity, one evaluates   
the emission of gravitational waves, and takes into account 
the radiation damping effect as a correction.
In the cosmological context this method has been frequently used 
for the estimate of the emission rate of radiations~\cite{VE1982,VEV1984}. 

One might anticipate that arbitrary accuracy will 
be obtained by the successive iteration of this scheme.    
One way to justify this procedure is 
to solve both the equation of motion for the brane and 
the bulk field equations simultaneously, and compare the result with 
that obtained by the successive approximation.  
Though this program is impossible to accomplish 
in general situations, there are actually several exactly solvable 
cases. 
Making use of the high symmetry of the background geometries, 
the full set of perturbation equations of a self-gravitating 
vacuum brane in $4$-dimensional spacetime with infinitesimal width 
have been solved~\cite{KIF1994,TS1997,II1997}.  
The unperturbed configurations considered in these references 
are composed of maximally symmetric 4-dimensional spacetime and 
a vacuum de Sitter 2-brane.  
It has been shown that, once the gravitational backreaction is turned on, 
a vacuum brane loses its own dynamical degrees of freedom,  
and the perturbed motion is possible only while incidental gravitational 
waves come across the brane. 
Namely, the conclusion of those references was that 
a gravitating brane is not able to oscillate freely. 
On the other hand, when we consider a test non-gravitating brane, 
a small deformation of a brane in a maximally symmetric 
configuration is described 
as a scalar field on the unperturbed worldsheet, which 
obeys the Klein-Gordon equation with negative mass 
squared~\cite{GV1991,Guven1993}. 
The degrees of freedom corresponding to this scalar field 
have not been identified in the former description in which 
the effect of self-gravity is fully taken into account.  
In this sense, the physical picture described by 
the dynamics of a self-gravitating brane is quite different 
from that by a non-gravitating one. 

In this paper, we shall revisit the perturbation dynamics of a brane 
coupled to the bulk gravity, aiming at filling the gap between 
these two ways of describing the brane dynamics. 
Just for simplicity, 
we shall restrict our attention to a 
vacuum brane of co-dimension one.   

In Sec.~\ref{sect:non-gravitating}, we remind the description by 
means of a non-gravitating brane. In the first part of this 
paper~(Secs.~\ref{sect:background} - \ref{sect:selfperturbation}), 
after briefly summarizing the unperturbed 
background geometries in Sec.~\ref{sect:background}, 
we provide the general formula for bulk and brane perturbations in 
Secs.~\ref{sect:bulk-perturbation} and~\ref{sect:selfperturbation}, 
respectively. The formalism developed there 
is given in terms of the master scalar variable 
introduced in Refs.~\cite{KIS,Muko2000} and its variant,     
which considerably simplifies the perturbation equations. 
In Sec.~\ref{sect:bulk-brane-interaction}, we discuss 
the interaction between the motion of a gravitating brane 
and bulk metric perturbations, deriving the boundary conditions for 
the bulk perturbations along the brane from the perturbed 
junction conditions. 
We also derive the equation of motion for a gravitating brane, 
and compare it with the equation for a non-gravitating brane 
in Sec.~\ref{subsect:brane-pert. and master variable}.  
In the second 
part~(Secs.~\ref{sect:fill-the-gap} and \ref{sect:green-function}), 
we address the issue that in which situation   
the test non-gravitating brane treatment can 
well approximate the dynamics of a self-gravitating brane. 
We propose a weak backreaction condition 
in Sec.~\ref{sect:weakbackreaction}. 
In the succeeding subsections, we examine perturbations of a de Sitter brane 
embedded in a de Sitter bulk spacetime~(\dsds(n+1/n+2)) 
to show that when the weak backreaction condition holds 
the dynamics of a self-gravitating brane reproduces 
the picture described by a non-gravitating brane. 
In Sec.~\ref{sect:green-function} 
we complete our explanation of the correspondence between 
the two different ways of the description of brane dynamics,  
by constructing the retarded Green's function for the 
perturbation of a self-gravitating brane, 
and by identifying a pole in the expression of the Green's 
function with the perturbations corresponding to brane fluctuations.  
We summarize our results in Sect.~\ref{sect:summary}.   
In appendix~B, we explicitly construct the global solutions 
for the perturbations in some exactly solvable cases \dsds(3,5/4,6).

\section{Dynamics of a non-gravitating vacuum brane} 
\label{sect:non-gravitating}

First, we recapitulate an approximate treatment given 
by considering a non-gravitating vacuum brane on a fixed curved 
background~\cite{GV1991,Guven1993}. 
We are concerned with an $n$-brane $\Sigma$ embedded in 
$(n+2)$-dimensional spacetime~$(\tilde \cM,\tilde g_{MN})$. 
We use $\tilde x^M$ for the coordinates of the bulk spacetime, 
while $x^\mu$ for the coordinates on the brane. 
The metric induced on the brane is given by 
$q_{\mu \nu}= \tilde{g}_{MN}
\partial_\mu \tilde{x}^M \partial_\nu \tilde{x}^N$. 
The brane $\Sigma$ divides the ambient manifold $\tilde{\cal M}$ 
into two parts, which we respectively call 
$\tilde{\cal M}_+$ and $\tilde{\cal M}_-$. 
Dynamics of a non-gravitating brane follows the equations of 
motion derived from Nambu-Goto type action 
\begin{equation}
   S = -\sigma \int_\Sigma d^{n+1}x \sqrt{-\det(q_{\mu \nu})} 
  +\Delta\rho\int_{\tilde{\cM}_-} d^{n+2}\tilde{x}\sqrt{-\det(\tilde{g}_{MN})} \,,  
\label{action:NG}  
\end{equation}
where $\sigma$ is the brane tension, 
and $\Delta\rho$ is the difference of the values of the 
vacuum energy density between 
$\tilde{\cM}_+$ and $\tilde{\cM}_-$. 
We assume that both $\sigma$ and $\Delta\rho$ are constant throughout this paper.   
In the non-gravitating brane approximation, 
we neglect the effects of gravity caused by $\Delta\rho$  
as well as those by $\sigma$.  
Starting with this action, and taking the variation with respect 
to a small deviation in the brane configuration $\tilde{x}^M(x^\mu)$,  
we have the equation of motion for a non-gravitating brane as 
\begin{equation}
   K^\mu{}_\mu = - \frac{\Delta\rho}{\sigma} \,.    
\label{eom:trK}
\end{equation}
Here the extrinsic curvature is defined by $K_{\mu \nu}:= - \tilde{g}_{LN}
\partial_\mu \tilde{x}^M \partial_\nu \tilde{x}^L \tilde{\nabla}_M n^N$,  
with $n^N$ being a unit vector normal to $\Sigma$ pointing toward 
$\tilde{\cal M}_+$, and $\tilde{\nabla}_M$ the covariant derivative 
with respect to the bulk metric $\tilde{g}_{MN}$. 

The perturbations of a brane configuration, 
$\tilde{x}^M \rightarrow \tilde{x}^M + \delta \tilde{x}^M $,   
can be decomposed as 
\begin{equation}
    \delta \tilde{x}^M = Z_\para^M + Z_\perp n^M \,,   
\label{deform:brane-configuration}
\end{equation} 
where $Z_\para^M$ is a vector and $Z_\perp$ is a scalar 
living on the unperturbed brane worldsheet. 
The deformations tangential to the brane $Z_\para^M $ 
are nothing but 
worldsheet diffeomorphisms, and 
the only transverse deformation $Z_\perp n^M$ is 
physically relevant~\cite{GV1991,Guven1993}. 
Namely, the perturbations of the brane configuration are 
described by a single scalar field $Z_\perp$. 
The deformation of the brane on a fixed bulk generates 
changes in the induced metric and also in the extrinsic curvature 
as
\begin{equation}
  \delta q_{\mu \nu} = - 2 K_{\mu \nu} Z_\perp \,, \quad 
  \delta K_{\mu \nu} = \nabla_\mu \nabla_\nu Z_\perp 
              + \left(
                      \tilde{R}_{\perp \mu \perp \nu} 
                      - K^\sigma{}_\mu K_{\sigma \nu} 
                \right) Z_\perp \,, 
\end{equation} 
where 
${\nabla}_\mu$ is the covariant derivative with respect to 
the unperturbed background induced metric $q_{\mu \nu}$, and 
$\tilde{R}_{\perp \mu \perp \nu} 
 := n^M n^N \partial_\mu \tilde{x}^L \partial_\nu \tilde{x}^K 
    \tilde{R}_{M L N K}$ 
with $\tilde{R}^M{}_{L N K}$ being the background bulk curvature tensor 
evaluated at the location of the brane. 
{}From the condition that $\delta K^\mu{}_\mu=0$, we obtain 
\begin{equation}
  \left(\Box_{(n+1)} - m^2 \right) Z_\perp = 0 \,,     
\label{eom:non-grav-brane}
\end{equation}
where $\Box_{(n+1)}:= \nabla^\mu \nabla_\mu$, and the mass-squared 
is given by 
\begin{equation}
 m^2 = - \tilde{R}_{\perp \mu \perp \nu} q^{\mu \nu}
                      - K^\nu{}_\mu K^\mu{}_\nu 
     = - \frac{{}^{(n+1)}\!R}{n} \,,    
\label{mass-squared} 
\end{equation} 
with ${}^{(n+1)}\!R$ being the scalar curvature of the unperturbed 
worldsheet. 
Thus the brane perturbation obeys Klein-Gordon type wave equation defined 
on the unperturbed worldsheet. 
{}For dS-branes, the behavior of the solutions of 
Eq.~(\ref{eom:non-grav-brane}) has been studied in detail 
in Ref.~\cite{GV1991}.

\section{Background geometry}
\label{sect:background}
Next, we discuss the dynamics of a self-gravitating vacuum brane with co-dimension one. 
We assume that the bulk spacetime of the 
unperturbed background is 
maximally symmetric. Namely, it is given by 
$(n+2)$-dimensional de Sitter (\ds$_{n+2}$), Minkowski 
(\min$_{n+2}$), or anti-de Sitter (\ads$_{n+2}$) space,  
depending on the value of the bulk cosmological constant $\lambda$.  
As we will see, the unperturbed geometry induced on the brane 
also becomes maximally symmetric. 
Such solutions are well known, but we briefly recapitulate 
them in order to establish our notation.  

Anticipating future extension, 
we begin with more general framework merely assuming that 
the bulk metric admits the isometry of (not $n+2$ but) 
$n$-dimensional maximally 
symmetric space $\cK^n$ with constant curvature $K$. 
The metric takes the form of
\begin{equation}
  d\tilde{s}_{(n+2)}^2 = \tilde{g}_{MN}d\tilde{x}^M d\tilde{x}^N 
               = g_{ab}(y) dy^a dy^b + r^2(y) d\sigma_{(n,K)}^2\,, 
\label{metric:bulk}  
\end{equation}
where $d\sigma_{(n,K)}^2 = \gamma_{ij} dx^idx^j$ is the metric of $\cK^n$.  
We assume that the brane admits the same isometry as $\cK^n$ so that 
the induced metric $q_{\mu\nu}$ takes the form of 
\begin{equation} 
  q_{\mu \nu}dx^\mu dx^\nu 
 = \alb^2 \left\{ -d\tau^2 + A^2(\tau)\,d\sigma_{(n,K)}^2 \right\} \,,  
\label{metric:induce}
\end{equation} 
where $\tau$ is a proper time normalized by a constant $\alb$, 
and $\alb A(\tau)$ is the restriction of the function $r(y)$ 
on $\Sigma$, i.e., $\alb A(\tau)=r(y)|_{\Sigma}$.   
We introduced $\alb$ just for convenience, and we specify its 
value later. 

The Israel's junction condition implies 
that the difference of the values of the extrinsic curvature 
evaluated on both sides of $\Sigma$ is related to 
the intrinsic energy-momentum tensor $T_{\mu \nu}$ as 
\begin{equation}
 \left[K_{\mu\nu}\right] 
   := K_{+\mu\nu}  - K_{-\mu \nu} 
   = \tilde{\kappa}^2 \left( T_{\mu\nu} - \frac{1}{n}T q_{\mu\nu} \right) \,,  
\label{jc:extrinsic-curvature}  
\end{equation}
where $\tilde{\kappa}^2$ is the $(n+2)$-dimensional 
gravitational constant.  
This condition comes from the $(\mu,\nu)$-components 
of the Einstein equations.  
Hereafter, quantities with subscripts $\pm$ denote 
their value evaluated on the respective sides 
$\tilde{\cal M}_\pm$.
We denote the difference of 
the values of a variable $Q$ evaluated on both sides 
by $[Q]:= Q_+ - Q_-$ as used above, while the averaged value by 
\begin{equation}
  \overline{Q} := \half ( Q_+ + Q_- ) \,. 
\end{equation}

On the assumption that the intrinsic energy-momentum tensor of a vacuum brane 
is given by $T_{\mu\nu} = - \sigma q_{\mu\nu}$, 
the junction condition (\ref{jc:extrinsic-curvature}) becomes 
\begin{eqnarray}
 \left[ {D_\perp r \over r} \right] &=& 
 - \frac{\tilde{\kappa}^2 \sigma}{n} \,,  
\label{jc:background-1}
\end{eqnarray}  
where $D_\perp := n^aD_a$ and 
$D_a$ is the covariant derivative with respect to the orbit space 
metric $g_{ab}$. 
In a similar manner we use an abbreviation 
$D_\para$ to represent the covariant derivative projected 
in the direction tangential to the brane. 
Substituting Eq.~(\ref{jc:extrinsic-curvature}) into 
the Hamiltonian and the momentum constrains,   
we also obtain 
\begin{eqnarray}
 T^{\mu \nu}\overline{K}_{\mu \nu} &=& 
 \frac{n(n+1)}{2\tilde{\kappa}^2}\left[ \lambda \right] \,,
\label{jc:average}
\\
 \nabla_\nu T^\nu{}_\mu &=& 0 \,, 
\label{jc:conservation}
\end{eqnarray}  
where we assumed the $(n+2)$-dimensional vacuum Einstein equations 
with cosmological constant, 
$\tilde{G}_{MN} = \tilde{R}_{MN} - \half\tilde{R}\tilde{g}_{MN} 
  = - \half n(n+1)\lambda\tilde{g}_{MN}$ in the bulk. 
 
Now we focus on the case that $\tilde{\cal M}_+$ and $\tilde{\cal M}_-$ 
are both composed of locally maximally symmetric spacetime. 
In this case, the metric function $r(y)$ satisfies 
\begin{equation}
-(D_\para r)^2 +(D_\perp r)^2 -K+ \lambda r^2=0\,,
\label{req}
\end{equation}
in respective sides, $\tilde{\cal M}_+$ and $\tilde{\cal M}_-$.  
Then, the equation of motion for the brane 
(\ref{jc:background-1}) reduces to 
\begin{equation}
 \frac{K+ (D_\para r)^2}{r^2} 
 = {
     n^4(\lambda_+ - \lambda_-)^2 + \tilde{\kappa}^8\sigma^4 
         + 2n^2 \tilde{\kappa}^4 \sigma^2(\lambda_+ + \lambda_-)
   \over 
     4n^2\tilde{\kappa}^4 \sigma^2 
   } =: \frac{\eta}{\al_B^2} \,,    
\label{eom:background-brane}
\end{equation} 
where $\eta$ takes a value $-1, 0$, or $1$. 
When $\eta\ne 0$, we choose the value of $\alpha_B$ 
so as to satisfy this condition. 
In the case of $\eta=0$, we simply set $\alpha_B=1$. 

Since $r(y)|_{\Sigma}=\alb A(\tau)$ and $D_\para=
\alb^{-1}(\partial_{\tau})^a D_a$ on the brane, 
Eq.~(\ref{eom:background-brane}) reduces to $K+(\partial_\tau A)^2=\eta A^2$. 
We can immediately solve this equation to find 
that the brane has a locally maximally symmetric geometry. 
In all possible combinations of $K$ and $\eta$, 
we can introduce Gaussian normal coordinates 
as 
\begin{equation} 
d\tilde{s}_{(n+2)}^2 = d\chi^2 + q_{\mu \nu}dx^\mu dx^\nu 
                   = d\chi^2 + \alpha^2(\chi) 
                \left\{ - d\tau^2 + A^2(\tau) d\sigma_{(n,K)}^2 \right\} \,,   
\label{chart:gaussian}  
\end{equation} 
for the bulk geometry, and thus 
\begin{equation} 
  r(y^a) =  \alpha(\chi)A(\tau) \,. 
\end{equation} 
We specify the position of the brane at $\chi=\chi_B$, and therefore  
$\alb =\al(\chi_B)$. 
In these coordinates, the unit normal vector $n^a$ to the brane 
is naturally extended into the bulk by $(\partial_\chi)^a$ 
as the unit normal to $\chi = const.$ 
hypersurface. Correspondingly, the extension of 
the unit tangent is given by $\alpha(\chi)^{-1}(\partial_{\tau})^a$.  
{}From Eq.~(\ref{req}), we also have 
\begin{equation}
  \left({D_\perp r\over r}\right)^2=-\lambda+{\eta\over \alpha^2} \,,   
\label{alphaeq}
\end{equation}
or more explicitly
$(\partial_\chi \alpha)^2=-\lambda\alpha^2+\eta$.

The unperturbed background geometry is summarized as follows.  
\begin{itemize}
\item 
$\eta=-1$; 
{\bf \adsads(n+1/n+2):} 
In this case, $K$ must be $-1$ for the existence of  a solution of 
Eq.~(\ref{eom:background-brane}). 
The solution 
\begin{equation} 
     A(\tau) =  \cos \tau \,, \quad  (\mbox{for}~ K = -1) \,, 
\end{equation}
indicates that the brane is adS space. 
In order that Eq.~(\ref{alphaeq}) has a solution, 
$\lambda$ must be negative. 
For convenience, we introduce the bulk curvature radius $\ell$ 
defined by $|\lambda|=\ell^{-2}$. 
Then, the solution on each
side $\tilde{\cal M}_+$ or $\tilde{\cal M}_-$ is given by 
\begin{equation}
    \alpha(\chi) 
                 = \ell \cosh \Frac(\chi /\ell) \,, \,\, 
        \left( \mbox{for}~ \lambda < 0        \right) \,,        
\end{equation} 
up to a shift of the origin of the $\chi$-coordinate. 
The embedding bulk is adS space.   

\item 
$\eta=0$; 
{\bf \minads(n+1/n+2) or $\min_{n+2}$:} 
There are two cases $K=0$, and $-1$. 
In both cases, the solutions 
\begin{equation}
  A = constant , \,\,(\mbox{for}~ K =0) \,, \quad 
  A(\tau) = \tau , \,\, (\mbox{for}~ K = -1) \,, 
\end{equation}
show that the brane geometry is given by Minkowski space. 
The warp factor is given by 
\begin{equation}
    \alpha(\chi) 
                = \ell \exp \Frac( \chi / \ell ) \,, \,\, 
        \left( \mbox{for}~ \lambda < 0  \right) \,,  \quad      
      \alpha(\chi)=\alpha_B\,,\,\,         
        \left( \mbox{for}~ \lambda = 0  \right) \,. 
\end{equation} 
The bulk is therefore adS, or Minkowski space.

\item 
$\eta=1$; 
{\bf $\ds_{n+1}$-brane $\subset$ $\ds_{n+2}$-, $\min_{n+2}$-, 
or $\ads_{n+2}$-bulk:} 
The scale factor on the brane is given by   
\begin{equation}
  A(\tau) = \sinh\tau\,, \,\,(\mbox{for}~ K =-1) \,, \quad 
  A(\tau) =\exp \tau\,, \,\,(\mbox{for}~ K =0) \,, \quad 
  A(\tau) =\cosh\tau\,, \,\,(\mbox{for}~ K =1) \,.
\end{equation}
The warp factor is given by 
\begin{equation}
   \alpha(\chi) = \ell \sinh \Frac(\chi/\ell) \,, \,\, 
        \left( \mbox{for}~ \lambda < 0  \right) \,,        
  \quad 
    \alpha(\chi) = \chi \,, \,\, 
        \left( \mbox{for}~ \lambda = 0  \right) \,,\quad        
   \alpha(\chi) = \ell \cos \Frac(\chi/\ell) \,, \,\, 
        \left( \mbox{for}~ \lambda > 0  \right) \,.         
\end{equation}  
This means that de Sitter brane can be embedded 
into an ambient spacetime consisting of 
any combination of the two maximally symmetric spacetimes. 
\end{itemize}

\section{Bulk-perturbation}  
\label{sect:bulk-perturbation}

In this section, we provide the bulk perturbation equations, 
following the gauge-invariant formalism for perturbations developed 
in Ref.~\cite{KIS}. 
Perturbations of the present bulk geometry can be decomposed 
into the tensor-type, the vector-type, and the scalar-type components 
classified by the number of tensor indices tangent to 
the $n$-dimensional invariant space~$(\cK^n,d\sigma_{(n,K)}^2)$. 
The Einstein equations for each type of perturbations decouple 
from the others. 
We focus on the scalar-type perturbation 
since the displacement of a brane does not couple to the
others as was shown in Ref.~\cite{KIS}.  
We introduce the gauge-invariant variables for the scalar-type 
bulk perturbation.
It turns out that these gauge-invariant variables can be described 
in terms of a single scalar master variable, and the perturbation equations 
reduce to a single equation for this master variable. 
  
The scalar harmonics $\Bbb S$ defined on $(\cK^n,d\sigma_{(n,K)}^2)$ is 
a function which satisfies 
\begin{equation}
 (\hat{\triangle} + k^2 ){\Bbb S}_{\bf k} = 0 \,, 
\end{equation}
where $\hat \triangle:= \hat D_i \hat D^i$, and 
$\hat D_i$ is the covariant derivative 
on~$(\cK^n,d\sigma_{(n,K)}^2)$. 
The scalar-type harmonic vector and tensor 
are constructed from $\Bbb S_{\bf k} $ as 
\begin{equation}
 {\Bbb S}_{{\bf k}\,i} = -\frac{1}{k}\hat{D}_i{\Bbb S_{\bf k}} \,, \quad 
 {\Bbb S}_{{\bf k}\, ij} = \frac{1}{k^2}\hat{D}_i\hat{D}_j {\Bbb
       S_{\bf k}} + \frac{1}{n}\gamma_{ij}{\Bbb S_{\bf k}} \,. 
\end{equation}  
The metric perturbations of scalar-type 
are expanded as 
\begin{equation} 
  h_{ab} = \sum_{\bf k}
           f^{\bf k}_{ab}{\Bbb S}_{\bf k} \,, \quad 
  h_{ai} = \sum_{\bf k}
           r f_{a}^{\bf k}{\Bbb S}_{{\bf k}\,i} \,, \quad 
  h_{ij} = \sum_{\bf k}
           2r^2 \left(H_L^{\bf k} \gamma_{ij}{\Bbb S}_{\bf k} 
           + H_T^{\bf k} {\Bbb S}_{{\bf k}\, ij} \right) \,,     
\end{equation} 
where the coefficients $f_{ab}$, $f_{a}$, $H_L$, and $H_T$ are functions 
on the $2$-dimensional orbit space $(\cN^2,g_{ab})$.  
In the following we abbreviate the indices ${\bf k}$ labeling the
eigenvalues of the harmonics and the summation over them. 

Under the action of an infinitesimal coordinate transformation 
$\tilde{x}^M \to \tilde{x}^M + \xi^M $, the metric perturbations 
transform as 
\begin{equation}
 h_{MN} \rightarrow h_{MN} +\bar{\delta}h_{MN} 
           = h_{MN} - \tilde{\nabla}_M \xi_N - \tilde{\nabla}_N \xi_N \,. 
\end{equation} 
The generator of infinitesimal gauge-transformation for the scalar-type 
perturbation is expanded as 
\begin{equation}
   \xi_a = T_a {\Bbb S} \,, \quad \xi_i = rL {\Bbb S}_i \,. 
\label{generator-gauge}
\end{equation}
Then, the metric components, 
$f_{ab}$, $f_{ai}$, $H_L$, and $H_T$, transform as 
\begin{eqnarray} 
 && \bar\delta f_{ab}=-D_a T_b -D_b T_a \,, 
\\
 && \bar\delta f_a=-rD_a\Frac(L/r)+\frac{k}{r}T_a \,,
\\
 && \bar\delta H_L=-\frac{k}{nr}L-\frac{D^ar}{r}T_a \,,
\\
 && \bar\delta H_T=\frac{k}{r}L \,.  
\end{eqnarray} 
{}From these transformation laws, 
we find gauge-invariant combinations of the variables as 
\begin{equation}
   F := H_L + \frac{1}{n}H_T + \frac{D_a r}{r}X^a \,, \quad 
   F_{ab} := f_{ab} + D_a X_b + D_b X_a \,,  
\end{equation}  
where a vector in ${\cal N}^2$ was introduced by
\begin{equation}
    X_a := \frac{r}{k}\left( f_a + \frac{r}{k}D_a H_T \right) \,, 
\label{def:Xa}
\end{equation}  
which transforms as $\bar\delta X_a=T_a$. 
Note that ${\Bbb S}_i$ and ${\Bbb S}_{ij}$ vanish identically for $k=0$ mode,  
and ${\Bbb S}_{ij}$ also vanishes for $k^2=nK$ mode. 
The gauge invariant variables which contain expansion coefficients 
of the harmonics at these special values of $k^2$ do not exist 
from the beginning. 
Hence, the equations containing these variables are no longer valid. 
In what follows, we do not consider $k^2(k^2-nK)= 0$ modes.  

As shown in Ref.~\cite{KIS}, the bulk Einstein equations for the scalar-type 
perturbation reduce to a set of equations for the gauge-invariant variables 
$F$ and $F_{ab}$ (See appendix~A.). 
Among them, we can identify the constraint equations 
\begin{eqnarray}
   F^a{}_a + 2(n-2)F = 0 \,, \quad 
  D_b (r^{n-2} F^b{}_a )= 2D_a(r^{n-2}F )\,. 
\label{bulk:constraints}
\end{eqnarray} 
From these constraint equations, 
we can show that there exists a master variable $\tilde\Omega$ in terms of 
which the gauge-invariant variables above can be expressed as~\cite{KIS} 
\begin{eqnarray}
2n r^{n-3} F &=& \left\{ \Box + 2 \frac{D^c r}{r}D_c
                 \right\}\tilde\Omega \,,  
\label{express:FbyOmega} 
\\
 r^{n-3} \{ F_{ab} + 2(n-1)F g_{ab} \} 
   &=& \left\{ D_a D_b 
             + \frac{D_a r}{r}D_b+\frac{D_b r}{r}D_a 
       \right\}\tilde\Omega \,, 
\label{express:FabbyOmega} 
\end{eqnarray}  
where $\Box := D^aD_a$. 
{}From the evolution equations of the Einstein equations, the equation
for the master variable $\tilde\Omega$ reduces to  
\begin{equation}
\left\{\Box - (n-2) \frac{D^c r}{r} D_c - \frac{k^2}{r^2} \right\}  
 \tilde\Omega = 0 \,.   
\label{eq:reduced-master}
\end{equation}
In the Gaussian normal coordinates~(\ref{chart:gaussian}), 
the master equation above is expressed as 
\begin{eqnarray}
&&  
  \left\{ 
     A^{n-2} \prtl_\tau A^{-(n-2)} \prtl_\tau + \frac{k^2}{A^2} 
     + p^2 + \eta\nu^2 
  \right\} \cT_p(\tau) = 0 \,, 
\label{eq:T}  
\\
&&  
 \left\{ 
        \al^{n-3}\prtl_\chi \al^{-(n-3)} \prtl_\chi 
        + \frac{1}{\al^2} \left( p^2 + \eta \nu^2 \right) 
 \right\} \cR_p(\chi) = 0 \,, 
\label{eq:R}
\end{eqnarray}
where we assumed a separable form of 
the solution as $\tilde\Omega = \cT(\tau) \cR(\chi)$, and 
introduced $p^2$ as a separation constant.  
Here, we also defined 
$$
 \nu := \frac{n-2}{2}\,.
$$  
The bulk Einstein equations for scalar perturbations and 
the derivation of the master equation are briefly summarized in 
appendix~A.

\section{Perturbation of a self-gravitating brane}
\label{sect:selfperturbation}

In this section, analyzing the junction condition, 
we discuss the interrelation between the bulk metric 
perturbations and the displacement of the brane.   
The junction condition provides the boundary condition for 
the bulk metric perturbations on the brane, 
which we need in finding the global solutions of the bulk perturbations.   
Simultaneously, we obtain an expression 
for the gauge-invariant brane displacement solely written in terms of 
the master variable from the junction condition. 
Hence, we find that there are no dynamical degrees of freedom 
for the brane fluctuations independent of 
the bulk metric perturbations. 
Combined with the Hamiltonian constraint, the junction condition also 
provides the equation for the displacement 
of a self-gravitating brane, which is to be compared with that 
for a non-gravitating brane obtained from the Nambu-Goto action.

\subsection{Coupling of bulk and brane-perturbations} 
\label{sect:bulk-brane-interaction}

To treat perturbations of a self-gravitating brane 
in a gauge invariant manner,  
the bulk metric perturbations  
have to be taken into account as well as the perturbations 
of the brane configuration. 
The perturbation of the brane configuration, $Z_\perp$,  
transforms in the same way as 
a combination of the bulk metric perturbations $X_\perp$
under the infinitesimal coordinate-transformation~(\ref{generator-gauge}); 
$\bar{\delta}Z_\perp=\bar{\delta}X_\perp = T_\perp$. 
Thus we immediately find a gauge-invariant variable corresponding to 
the displacement of the brane as 
\begin{equation}
   Y_\perp := Z_\perp - X_\perp \,. 
\end{equation}  
Since $Y_\para := Z_\para - X_\para$  
is of no physical relevance as mentioned before,   
we simply set $Y_\para$ to zero in what follows.  
   
Taking the bulk perturbation into account, we find that 
the perturbations of the induced metric and 
the extrinsic curvature are given by  
\begin{eqnarray} 
  \delta q_{\mu \nu} &=& - 2 K_{\mu \nu} Z_\perp + h_{\mu \nu} \,, 
\label{del:qmunu}
\\ 
  \delta K_{\mu \nu} &=& \nabla_\mu \nabla_\nu Z_\perp 
              + \left(
                      \tilde{R}_{\perp \mu \perp \nu} 
                      - K^\sigma{}_\mu K_{\sigma \nu} \right) Z_\perp 
\non \\
      && \quad    
          + \half n^a \left(\tilde{\nabla}{}_\mu h_{a \nu} 
                         + \tilde{\nabla}_\nu h_{a \mu}  
                         - \tilde{\nabla}_a h_{\mu \nu} 
                   \right)
          + \half h_{\perp \perp} K_{\mu \nu} \,.  
\label{del:Kmunu}
\end{eqnarray}
Since $\delta K^\mu{}_\nu$ is gauge invariant, 
we can express it 
solely in terms of the gauge-invariant variables $F$, $F_{ab}$, 
and $Y_\perp$ as~\cite{KIS},  
\begin{eqnarray} 
\delta K^\tau{}_\tau &=&
   \left\{
         -\half D_\para F_{\perp \para}+\half n_aD_b F^{ab} 
         -\half D_\perp F^a{}_a + \half K^\tau{}_\tau F_{\para\para} 
         - D_\para^2 Y_\perp  
          + \frac{{}^{(n+1)}\!R }{n(n+1)} Y_\perp 
   \right\}{\Bbb S} \,, 
\label{pert-K-tt} 
\\
 \delta K^\tau{}_i &=& 
   k\left\{  
        \half F_{\perp\para} 
        + r D_\para \Frac(Y_\perp/ r)
    \right\} {\Bbb S}_i \,, 
\label{pert-K-ti}
\\
\delta K^i{}_j &=& 
    \left\{
          - D_\perp F - \frac{ D_\para r}{r} F_{\perp\para}
          +\half \frac{D_\perp r}{r} F_{\perp \perp} 
          - \left( \frac{D_\para r}{r} D_\para 
                   + \frac{k^2}{n r^2} - \frac{{}^{(n+1)}\!R }{n(n+1)}  
            \right) Y_\perp 
    \right\}{\Bbb S}\, \delta^i{}_j 
    +\frac{k^2}{r^2}Y_\perp {\Bbb S}^i{}_j \,. 
\label{pert-K-ij} 
\end{eqnarray}  

Let us derive the boundary conditions for the master variable 
at the location of the brane. 
The perturbation of the induced metric can be made continuous 
across the brane 
\begin{equation}
   \left[ \delta q_{\mu \nu} \right] = 0 \,,  
\label{q:contiuous}
\end{equation}
by appropriately choosing the coordinates tangential to the brane  
because the intrinsic geometry on
both sides of the brane must be identical.  
{}From the junction condition~(\ref{jc:extrinsic-curvature}), we have  
\begin{equation}
   [ \delta K^\mu {}_\nu ] = 0 \,.   
\end{equation} 
For further computation, it is convenient to use the gauge $Z_\perp =0$ so that 
$X_\perp = -Y_\perp $ on the brane. 
Then, from the metric continuity~(\ref{q:contiuous}) we have 
\begin{equation}
 \left[ F \right] = - \left[ \frac{D_\perp r}{r}Y_\perp \right] \,,  
\label{FX}
\end{equation}
and   
\begin{equation}
  \left[ F_{\para \para} \right] 
  = - 2 \left[K^\tau{}_\tau Y_\perp \right] \,,   
\label{FparaX} 
\end{equation}  
where we have used 
$K^\tau{}_\tau \equiv \al^{-1} (\partial_\tau)^a D_\para\, n_a$.   
Since $K^\tau{}_\tau = - r^{-1}D_\perp r$ for the present background geometry, 
we find 
\begin{equation}
  \left[ F_{\para \para} + 2F \right] = 0 \,.    
\label{jc:F_pparaF} 
\end{equation}  
{}From Eq.~(\ref{pert-K-ti}) and 
the component proportional to ${\Bbb S}^i{}_j$ in 
Eq.~(\ref{pert-K-ij}), we have   
\begin{eqnarray} 
 &&  \left[ F_{\para \perp} \right] = 0 \,,  
\label{jc:F-para-perp}  
\\
&& 
   \left[ Y_\perp \right] = 0 \,.   
\label{jc:Yperp}
\end{eqnarray}  
{}From the component proportional to 
${\Bbb S}\,\delta^i{}_j$ in Eq.~(\ref{pert-K-ij}), we obtain 
\begin{equation}
 2 \left[ D_\perp F \right] 
   = \left[\frac{D_\perp r}{r} F_{\perp \perp} \right] \,, 
\label{jc:Dr-Fpperp} 
\end{equation} 
where we have used the formula  
\begin{equation} 
  \frac{D_aD_b r}{r} = - \lambda \; g_{ab} \,,  
\end{equation} 
and Eq.~(\ref{alphaeq}). 
No new independent condition is obtained from 
Eq.~(\ref{pert-K-tt}). 
Note that the results we obtained are gauge independent, 
even though they are derived by using a specific gauge.

Now let us express these conditions~(\ref{jc:F_pparaF}), 
(\ref{jc:F-para-perp}), and (\ref{jc:Dr-Fpperp}) in terms of 
the master variable $\tilde\Omega$, 
using Eqs.~(\ref{express:FbyOmega}) and~(\ref{express:FabbyOmega}).  
Firstly, we rewrite Eq.~(\ref{jc:F_pparaF}) as 
\begin{eqnarray} 
 0  &=&   
    \left\{ 
           D_\para^2 - (n-2) \frac{D_\para r}{r} D_\para 
           + \frac{k^2}{r^2} 
    \right\}  \left[ \tilde\Omega \right] 
    + (n-1)  \left[ \Frac(D_\perp r/r) D_\perp \tilde\Omega\right] \,,  
\end{eqnarray}  
where we have used the formula (\ref{eom:background-brane}). 
Further, using the master equation~(\ref{eq:T}), we can write 
the condition above as 
\begin{eqnarray} 
(n-1)\left[\left( \frac{D_\perp r}{r} \right)D_\perp \tilde\Omega \right] 
 = \frac{1}{\alpha^2} \left( p^2 + \eta \nu^2 \right) 
   \left[\tilde\Omega \right] \,. 
\label{jc:1} 
\end{eqnarray}
Secondly, from the condition~(\ref{jc:F-para-perp}), we have  
\begin{equation}
D_\para \left(r \left[ D_\perp \tilde\Omega \right] \right) = 0 \,. 
\label{eq:Fparaperp}
\end{equation} 
Thirdly, we write down the condition~(\ref{jc:Dr-Fpperp}) explicitly. 
{}From Eqs.~(\ref{express:FbyOmega}) 
and (\ref{express:FabbyOmega}), we have  
\begin{eqnarray}
2r^{n-1} F 
 &=& \left\{ r(D_\perp r) D_\perp-r(D_\para r)D_\para +{k^2\over n} \right\}
  \tilde\Omega \,,\cr 
 r^{n-3}\left\{F_{\perp \perp}+2(n-1) F \right\}
 &=& \left( D_{\perp}^2 + 2\frac{D_\perp r}{r}D_\perp\right)
   \tilde\Omega \,. 
\end{eqnarray} 
Applying $D_\perp$ to the first equation, we obtain 
\begin{eqnarray}
2r^{n-1}\left\{ (n-1){D_\perp r\over r} F+ D_\perp F \right\} 
 &=& \Biggl\{ (D_\perp r)^2D_\perp  + r(D^2_\perp r)D_\perp 
        + r(D_\perp r) D^2_\perp
\cr &&\quad\quad
-(D_\perp r)(D_\para r)D_\para 
        -r(D_\para r)D_\perp D_\para +{k^2\over n}D_\perp\Biggr\} 
  \tilde\Omega \,. 
\end{eqnarray}
Using these relations, we can rewrite the condition~(\ref{jc:Dr-Fpperp}) as 
\begin{equation}
   \frac{k^2 - nK}{nr^2} 
     \left[ D_\perp \tilde\Omega \right] = 
   \Frac(D_\para r/r) 
  r^{-1}D_\para \left(r \left[ D_\perp \tilde\Omega \right] \right)
  \,, 
\label{eq:Dr-Fpperp-Omega}
\end{equation}
where we have used 
\begin{equation}
\left({D_\perp r\over r}\right)^2-{D_\perp^2 r\over r}
   ={\eta\over\alpha ^2} \,, 
\end{equation}
which follows from Eq.~(\ref{alphaeq}). 

Combining Eqs.~(\ref{eq:Fparaperp}) and
(\ref{eq:Dr-Fpperp-Omega}), 
we find 
\begin{equation}
   \left[ D_\perp \tilde\Omega \right] = 0 \,. 
\label{jc:2b} 
\end{equation} 
Thus, $D_\perp \tilde\Omega$ is continuous across the brane. 
Then, from Eq.~(\ref{jc:1}), we have 
\begin{equation}  
  \frac{1}{\mu} \left(p^2 + \eta\nu^2 \right) 
  \left[ \tilde\Omega \right] 
   = - 2(n-1)\al_B D_\perp \tilde\Omega \,, 
\label{jc:1a} 
\end{equation}  
where we have introduced 
\begin{equation}
   \mu := \frac{\tilde{\kappa}^2\sigma\al_B}{2n} 
        = - \frac{\alpha_B}{2} \left[ \frac{D_\perp r}{r} \right]\,.  
\label{def:mu}
\end{equation}
The Eqs.~(\ref{jc:2b}) and (\ref{jc:1a}) are the boundary 
conditions for the master variable imposed on the brane. 

We should note that the master variable $\tilde\Omega$ is not 
continuous at the location of the brane. 
We introduce a new master variable 
defined by 
\begin{equation} 
    \cW := \alpha^{-(n-3)} D_{\perp}\tilde\Omega = \cT_p(\tau) \cQ_p(\chi) \,
\end{equation}
where
\begin{equation}
   \cQ_p(\chi) 
       := \alpha^{-(n-3)} D_{\perp} \cR_p  \,. 
\label{def:Q}
\end{equation} 
{}From Eqs.~(\ref{eq:R}) and (\ref{jc:2b}), we find that a quantity $\cQ_p$ 
defined by (\ref{def:Q}) is continuous, and so is $\cW$. 
In the bulk, $\cQ_p$ satisfies the equation 
\begin{equation}  
    \left\{
         \alpha^{-(n-1)}\prtl_\chi \alpha^{n-1} \prtl_\chi 
         + \frac{1}{\alpha^2} 
           \left(p^2 + \eta\nu^2 \right) 
    \right\} \cQ_p = 0 \,.   
\label{eq:Q}
\end{equation} 
{}From Eqs.~(\ref{eq:R}), (\ref{def:Q}), and (\ref{eq:Q}),
we obtain
\begin{equation}
  \left( p^2 + \eta\nu^2 \right) 
  \left[\tilde\Omega\right] 
    = - \alpha^{n-1}_B \left[D_\perp \cW\right] \,. 
\label{rel:Omega-W}
\end{equation} 
Then, we can derive the junction conditions for $\cW$ from 
Eqs.~(\ref{jc:2b}) and (\ref{jc:1a}) as 
\begin{eqnarray}
 \left[\cW\right] &=& 0 \,, 
\label{jc:W1}
\\ 
 \left[D_\perp\cW\right]&=&\frac{2(n-1)}{\alpha_B}\mu \;\cW \,.  
\label{jc:W2} 
\end{eqnarray}   
Therefore $\cW$ obeys 
\begin{equation}
 \cL \cW:= \left\{ \alpha^{-(n-1)}\partial_\chi \alpha^{n-1} \partial_\chi
      -{2(n-1)\over \alpha_B} \mu \, \delta(\chi-\chi_B) 
      - \frac{1}{\al^2} \left( A^{n-2}\prtl_\tau A^{-(n-2)} \prtl_\tau 
        + \frac{k^2}{A^2} \right) 
 \right\}   \cW =0. 
\label{realmaster}
\end{equation}

Although we exclusively use 
this variable $\cW$ or $\tilde\Omega$ in this paper,  
it will be also worth noting that in the bulk 
$\Phi:=\alpha^{-1} A^{-(n-1)}\cW$ 
satisfies the equation for a massive scalar field,  
$(\Box_{(n+2)} -n\lambda)\Phi=0$ with 
$\Box_{(n+2)}:=\tilde{\nabla}^M\tilde{\nabla}_M$, and 
$\Psi:=\alpha^{-(n-1)}\partial_\chi (\alpha^{n}\Phi)$ 
satisfies that for a massless scalar field, $\Box_{(n+2)}\Psi=0$. 
Furthermore, the junction conditions for $\Psi$ are simply given 
by $[\Psi]=[D_\perp \Psi]=0$.

\subsection{Equation of motion for $Y_\perp$}  
\label{subsect:brane-pert. and master variable}
From the condition (\ref{jc:average}),     
we can also derive an equation of motion for the brane fluctuations,   
which is analogous to that for a 
non-gravitating brane~(\ref{eom:non-grav-brane}).  
We will see, however, that the equation for the self-gravitating brane
differs in that it 
has a source term written in terms of the bulk 
metric perturbations. 

Let us derive the equation of motion for the gauge invariant 
displacement $Y_\perp$. 
{}From Eqs.~(\ref{pert-K-tt}) and (\ref{pert-K-ij}), the trace 
of $\delta K^\mu{}_\nu $ is computed as 
\begin{equation}
 {\delta K }^{\mu}{}_{\mu}
   = \left(\left\{ 
              - D_\para^2 - n\left(\frac{D_\para r}{r} \right)D_\para 
              - \frac{k^2}{r^2} - m^2 
       \right\} Y_\perp 
       + J \right){\Bbb S}\,,       
\end{equation} 
with
\begin{eqnarray}
   J  &:=&
          - \half r^{-(n+2)} D_\para (r^{n+2} {F}_{\para \perp}) 
          + \half \left(\frac{D_\perp r}{r} \right)F_{\perp \perp} 
          - D_\perp F \,, 
\label{J:wirttenbyF}
\end{eqnarray} 
where we have used the constraint equations~(\ref{bulk:constraints}). 
Then from the perturbation of the condition~(\ref{jc:average}),     
\begin{equation}
\delta  \overline{K}^{\mu}{}_\mu = 0 \,, 
\label{eom:delta-K}
\end{equation}  
we obtain the equation of motion 
for ${Y}_{\perp}$ as 
\begin{equation}
 \left( \Box_{(n+1)} - m^2 \right) {Y}_\perp + {J} = 0 \,,  
\label{eom:gravitatingbrane}
\end{equation}
with the mass-squared given by Eq.~(\ref{mass-squared}). 
Since there is no jump in $Y_\perp$ as shown in Eq.~(\ref{jc:Yperp}), 
we have replaced $\overline{Y}_\perp$ with $Y_\perp$. 
Similarly, we can show that 
the source term $J$ is also continuous across the brane 
using Eqs.~(\ref{jc:F-para-perp}) and~(\ref{jc:Dr-Fpperp}).  
Thus, we replace $\overline{J}$ with $J$. 
Alternative derivation of 
Eq.~(\ref{eom:gravitatingbrane}) is found in 
Ref.~\cite{II1999}.   
The appearance of a source term $J$ given by the bulk metric 
perturbations shows a clear difference from the equation for a non-gravitating 
brane~(\ref{eom:non-grav-brane}). 

More strikingly, $Y_\perp$ can be expressed solely in terms of 
the master variable. 
Using Eq.~(\ref{jc:background-1}), 
from Eqs.~(\ref{FX}), (\ref{FparaX}), 
and (\ref{jc:Yperp}), we immediately have 
\begin{equation}
    \left[F_{\para \para} - 2(n-2)F \right] 
    = - \frac{2(n-1)}{n} \tilde{\kappa}^2 \sigma Y_\perp \,.  
\end{equation} 
Then, writing $F_{\para \para}$ and $F$ in terms of $\tilde\Omega$, we obtain 
\begin{eqnarray} 
 \mu \:Y_\perp \!\!
  &=& - \frac{\al}{4(n-1)r^{n-3}}
         \left( 
               D_\para^2 + \frac{D_\para r}{r}D_\para 
               + \frac{k^2}{nr^2} 
         \right)\tilde\Omega \,.  
\label{eq:X-master-variable}  
\end{eqnarray}  
Hence, $Y_\perp$ is entirely expressed by the master variable $\tilde\Omega$. 
This equation indicates that the brane loses its own dynamical 
degrees of freedom, and a self-gravitating brane can oscillate only while 
bulk gravitational waves keep coming to it.   
Since this statement was first claimed in Ref.~\cite{KIF1994}, 
there has been a worry about the validity of the approximate description 
considering a non-gravitating brane, in which we assume 
the existence of the dynamical degrees of freedom corresponding 
to wall fluctuations {\it a priori}.

\section{Description of brane dynamics in the weak backreaction limit} 
\label{sect:fill-the-gap} 

In this section, we discuss the perturbative motion 
of a self-gravitating brane and compare it with that obtained by
considering a non-gravitating brane, 
using the formulas developed in the previous sections.  
We first propose a condition so that the brane fluctuations 
can be described with neglecting the gravitational backreaction 
of the bulk perturbations. 
We call it the weak backreaction condition. 
Whether or not one can take the weak backreaction limit 
depends on the system that we are concerned with.  
Next, analyzing \dsds(n+1/n+2) system for $n\geq 3$, 
we perturbatively obtain a global solution for the bulk metric perturbations 
which forms a discrete spectrum. 
We will see that the motion of a self-gravitating brane 
corresponding to this special mode reduces approximately to 
that of a non-gravitating brane. 
These observations will fill most of the gap 
between the two ways of describing the brane 
dynamics argued in the preceding section 
for $n\geq 3$. In the last subsection, we separately discuss 
the case with $n=2$, i.e., the case with 4-dimensional bulk.

\subsection{Weak backreaction limit} 
\label{sect:weakbackreaction}

Toward a resolution of the paradox mentioned in the 
preceding section, here 
a comment on the gauge dependence 
is in order.  
We can see from Eq.~(\ref{eq:X-master-variable}) that 
\begin{equation}
F_{ab},\, F \; \sim \; O(\tilde\Omega) \; \sim \; O(\mu Y_\perp) \,.
\label{Oder-relation:F-Omega-Y}
\end{equation}  
Let us assume $\mu$ is small. 
Then, it also follows by definition that 
\begin{equation}
    f_a, \, f_{ab}, \, H_L, \,H_T \;\sim\; O(X_\perp) \,. 
\end{equation}

When we choose the gauge such that $Z_\perp =0$, 
we have $Y_\perp = -X_\perp$, and therefore   
\begin{equation}
  f_a, \, f_{ab}, \, H_L, \, H_T \; 
  \sim \; O\left(Y_\perp \right) \,. 
\end{equation} 
In this gauge the dynamical degrees of freedom of the brane
fluctuations are completely transmuted into those of the 
bulk metric perturbations. 
On the other hand, if we choose the gauge such that $X_\perp =0$, 
we find $Y_\perp = Z_\perp$, $F_{ab} = f_{ab}$, 
$f_a = - k^{-1}{r}D_a H_T$ and $F=H_L+ \frac{1}{n}H_T$. 
Therefore
\begin{equation}
 f_a,\, f_{ab},\, H_L, \, H_T \; 
 \sim \; O\left( \tilde\Omega \right) \; 
 \sim \; O \left( \mu Y_\perp \right) \,. 
\end{equation}
Under this gauge choice, the brane fluctuations are mainly 
described by $Z_\perp$, 
and the perturbation of the bulk geometry stays small. 
This gauge choice provides a description relatively close to 
that obtained by considering a non-gravitating brane 
on a fixed background, if $\mu$ is sufficiently small. 

Based on the above observation, 
we propose that the condition that we can neglect the gravitational 
backreaction effects on the brane motion is specified by 
\begin{equation}
   \mu \ll 1 \,.   
\label{condi:weakbackreaction} 
\end{equation} 
We note that Eq.~(\ref{eom:background-brane}) can be expressed as 
\begin{equation}
  \frac{\eta}{\mu^2} = 1 + 
              \Frac(n^2 \left[\lambda\right] /\tilde{\kappa}^4 \sigma^2)^2  
              + {4n^2 \overline{\lambda}\over \tilde{\kappa}^4 \sigma^2} \,,  
\label{condi:mu=1+X+Y}
\end{equation} 
where $\eta=1,0,-1 $ represents the signature of the intrinsic 
curvature of the brane as before. 

{}For a \min-brane, Eq.~(\ref{condi:mu=1+X+Y}) becomes trivial if we
choose $\eta=0$. But by considering the limit of \min-brane from
$\eta=1$ or $-1$ case, one can see that 
$\mu^2$ goes to infinity in the \min-brane limit.  
Hence, the condition (\ref{condi:weakbackreaction}) cannot hold 
for \min-brane. 
{}For \ds-, and \ads-branes, 
the weak backreaction condition~(\ref{condi:weakbackreaction}) holds when 
\begin{eqnarray}
  \left| \left[\lambda \right]\right| &\gg& \tilde{\kappa}^4 \sigma^2 \,,
       \quad \mbox{or} 
  \quad \left|\overline{\lambda}\right| \gg \tilde{\kappa}^4 \sigma^2 \,. 
\label{condi:gg-1}
\end{eqnarray} 
The junction condition also gives 
\begin{equation}
 { [K^\mu{}_\mu] \over \overline{K}{}^\mu{}_\mu} 
  = - \frac{2\tilde{\kappa}^4 \sigma^2}{n^2\left[\lambda \right]} \,. 
\end{equation} 
Thus, the former case in Eq.~(\ref{condi:gg-1}) is 
equivalent to the condition 
that the gap between the values of the extrinsic curvature on both sides 
of the brane is sufficiently small.

One can also express Eq.~(\ref{eom:background-brane}) as 
\begin{equation}
  \frac{\eta}{\alpha^2 \lambda_+} 
   = 1
     + {\left(1-n^2[\lambda]/(\tilde{\kappa}^4 \sigma^2) \right)^2 
        \over 
       2n^2[\lambda]/(\tilde{k}^4\sigma^2)  
      +4n^2 \overline{\lambda}/(\tilde{k}^4\sigma^2)  
     } \,. 
\end{equation} 
Hence, apart from the exceptional case 
that $\left| \left[\lambda \right]\right|$ is 
extremely small compared with $\left|\overline{\lambda} \right|$, 
the weak backreaction condition~(\ref{condi:weakbackreaction}) 
implies that the curvature radius of the 
brane is much smaller than that of the bulk, 
i.e., $\ell_+ \gg \alpha$. 
In the exceptional case (
$\left|\overline{\lambda} \right|\gg \left| \left[\lambda
\right]\right|$ )
only the second possibility in Eq.~(\ref{condi:gg-1}) 
can be realized. 
In this case, we have $\alpha \sim \ell_- \sim \ell_+$, 
and the brane becomes an almost totally geodesic hypersurface. 
When we impose \z2-symmetry across the brane, 
we have $|[\lambda]|=0$. Hence, under \z2-symmetry
the weak backreaction condition is satisfied only through 
the second possibility in Eq.~(\ref{condi:gg-1}). 

\subsection{Global solutions in the weak backreaction limit} 

Here, focusing on \dsds(n+1/n+2) system with $n\geq 3$,   
we solve the lowest eigenmode of the bulk perturbations 
assuming the weak backreaction condition~(\ref{condi:weakbackreaction}). 
We find that there is at least one discrete eigenmode in this system. 
In the succeeding subsection, we will estimate 
the amplitude of the source term $J$ for this discrete mode, 
and we will show that the backreaction of the self-gravity on 
the brane motion is small. 
      
Just for convenience, we introduce a new coordinate $\zeta$ in place of 
$\chi$ by 
\begin{equation}
 \zeta(\chi):= \int \frac{d\chi}{\alpha(\chi)} 
 = \half \log \left\{\frac{1+\sin(\chi/\ell)}{1-\sin(\chi/\ell)} \right\} \,. 
\label{def:zeta}
\end{equation}
In terms of $\zeta$, the warp factor is given by
$\alpha=\ell\cos (\chi/\ell)= \ell/\cosh\zeta$. 
Here, as we have used the same functional form of 
the warp factor on both sides $\tilde{\cM}_\pm$, 
the $\chi$- and also $\zeta$-coordinates become 
discontinuous at the brane. 
Hence, in order to manifest this distinction, 
we associate $+$ or $-$ index with 
the coordinate $\chi$ and $\zeta$ if necessary.  
Using the new coordinate, we can rewrite Eq.~(\ref{eq:Q}) as 
\begin{equation}
    \left\{  
           \prtl_\zeta^2 + p^2 + \frac{\nu(\nu+1)}{\cosh^2\zeta} 
    \right\} \left( {\cQ_p \over \cosh^{\nu}\zeta}\right) = 0 \,.  
\label{eq:Z}
\end{equation} 
The solution that is regular also at $\zeta\to +\infty$ is given by 
\begin{equation} 
 \cQ_p = e^{\pi p/2}\Gamma(1+\nu) 
       (\cosh\zeta)^{\nu} P^{ip}_\nu (\tanh \zeta)\,.  
\quad 
\label{sol:W}
\end{equation} 
Once we notice the relation $P^{-\nu}_{\nu}(\tanh\zeta)=
(e^{-\nu\pi i}/\Gamma(1+\nu))(2\cosh\zeta)^{-\nu}$, 
it is easy to see that there is a solution regular at $\zeta\to -\infty$ 
for $p=i\nu$ in the absence of a brane. 
Since this solution is constant and hence nodeless, it is the lowest
eigenmode if it is normalizable.  
The standard normalization condition for discrete spectrum 
using the Hermitian measure is given by 
$\int d\zeta (\cosh \zeta)^{-2\nu}{\cQ}_p {\cQ}^*_{p'}=\delta_{p p'}$. 
Hence, it is actually normalizable for $n\geq 3$. 
We separately discuss $4$-dimensional bulk case $(n=2)$ in the succeeding
subsection. We will anticipate that the eigenvalue of 
this mode will shift from $p = i\nu$ once 
the gravity of the brane is turned on,  
but the corresponding mode itself will continue to exist. 
This shift of the eigenvalue can be calculated as follows. 
Setting $p^2 = - \nu^2+ \epsilon$, we seek such a shifted mode,  
and examine the behavior of $\cQ_p$ to the linear order in $\epsilon$. 
Let us introduce a function ${\cal E}(\chi)$ by 
\begin{equation}
 \cQ_p = \exp \left( \epsilon \int^\chi d{\chi'} \; 
       {\cal E}(\chi') \right) \,, 
\end{equation} 
where $\cal E$ is supposed not to be very large everywhere. 
Then, substituting this expression for $\cQ_p$ into 
Eq.~(\ref{eq:Q}), we have an equation for ${\cal E}$ 
to the lowest order in $\epsilon$ as 
\begin{equation}
 \al^{n-3} \partial_\chi \left( \al^{n-1} {\cal E}\right) + 1 = 0 \,. 
\end{equation} 
Integrating this, we find 
\begin{equation}  
 {D_\perp \cW_{(\pm)}\over \cW_{(\pm)}} = \epsilon \; {\cal E}_{(\pm)}
  (\chi_\pm) = - \epsilon \; \alpha^{1-n} 
                 \int^{\chi_\pm}_{\pm \pi \ell_\pm/2} 
                 d{\chi'_\pm }\alpha^{n-3}(\chi'_\pm ) \,,     
\end{equation}  
where ${\cal E}_{(\pm)}$ and hence $\cW_{(\pm)}$ satisfies the 
normalizability condition at $\zeta\to \pm \infty$.  
The jump in the logarithmic derivative of $\cW$ at 
the location of the brane is evaluated as 
\begin{eqnarray}
  \left[{D_\perp \cW \over \cW}\right] 
  &=& 
  \epsilon \; \alpha^{1-n} 
  \left\{ 
     \int^{\chi_{\raisebox{-2pt}{$\scriptscriptstyle -B$}}}_{-\pi \ell_- /2} 
     d\chi_- \alpha^{n-3}(\chi_-)
        + 
     \int^{\pi \ell_+ /2}_{\chi_{\raisebox{-2pt}{$\scriptscriptstyle +B$}}} 
     d\chi_+ \alpha^{n-3}(\chi_+)  
  \right\} 
\cr
  &\approx& 
  \epsilon \; \al^{1-n} 
    \int^{\pi \ell_+ /2}_{-\pi \ell_+ /2} d\chi_+ \al^{n-3}(\chi_+) 
\cr
 &=& 
    \sqrt{\pi} \epsilon \; \frac{\ell_+^{n-2}}{\al^{n-1}} 
    {\Gamma\left(\nu \right)\over \Gamma\left(\nu+\frac{1}{2}\right)} 
    \,, \quad (\mbox{for} \quad n \geq 3) \,,  
\end{eqnarray} 
where we have used the fact 
that $\ell_+ \approx \ell_-$ or $\alpha_B \ll \ell_+$ in 
the weak backreaction limit to obtain the approximation in the second line. 
Then, from the junction condition (\ref{jc:W2}),  
we have 
\begin{equation}
  \epsilon = \frac{4}{\sqrt{\pi}} 
  {\Gamma\left(\nu+\frac{3}{2}\right)\over \Gamma\left(\nu \right)} 
  \left(\frac{\alb}{\ell_+} \right)^{2\nu}\mu \,. 
\end{equation}
Hence, we find that the shifted eigenvalue is given by 
\begin{eqnarray}
   p &\approx& i \nu (1- \frac{\epsilon}{2\nu^2}) 
     \approx i  \nu 
             - i \frac{2}{\sqrt{\pi}}
  {\Gamma\left(\nu+\frac{3}{2}\right)\over \Gamma\left(\nu +1\right)} 
  \left(\frac{\alb}{\ell_+}\right)^{2\nu} \mu \,. 
\label{pole:n+2}
\end{eqnarray}

\subsection{Correspondence between two different descriptions 
of brane dynamics} 

Now we carefully examine the effective equation of motion 
for $Y_{\perp}$ given in Eq.~(\ref{eom:gravitatingbrane}) focusing 
on the specific mode obtained in the preceding subsection. 
Using the junction conditions for the master variable, 
we can express $J$ given in Eq.~(\ref{J:wirttenbyF}) 
in terms of the master variable $\tilde\Omega$. 
Inserting Eqs.~(\ref{express:FbyOmega}) and (\ref{express:FabbyOmega})  
into Eq.~(\ref{J:wirttenbyF}), 
with the aid of Eqs.~(\ref{jc:2b}) and (\ref{jc:1a}), we obtain 
after some calculation 
\begin{equation} 
 J = \frac{1}{\mu} \left( p^2 + \eta \nu^2 \right) 
     {\cO} \left[\tilde\Omega \right] \,,    
\label{eq:JandOmega}
\end{equation} 
with
\begin{equation}
 {\cO}:= \frac{1}{4(n-1)\alpha_B^{n} A^{n-3}} 
             \left\{ 
                   A^{-5} {\partial \over \partial \tau} 
                  A^5 {\partial \over \partial \tau}   
                   + \frac{1}{n} \frac{k^2}{A^2} 
                   + 4\left(\eta - \frac{K}{A^2} \right) 
             \right\} \,. 
\end{equation}  
This expression can be also obtained by directly acting the operator 
\begin{equation} 
   - D_\para^2 - n\Frac(D_\para r/r)D_\para - \frac{k^2}{r^2} - m^2\,,
\end{equation} 
on both sides of Eq.~(\ref{eq:X-master-variable}). 
By using Eqs.~(\ref{rel:Omega-W}) and (\ref{jc:W2}), $J$ can be 
also written by using $\cW$ as 
\begin{equation}
 J = -2(n-1)\alpha_B^{2\nu} {\cO} \cW \,.  
\end{equation}

The equation~(\ref{eq:JandOmega}) 
involves the eigenvalue $p^2$, which is determined 
only after solving the bulk metric perturbations. 
Hence, Eq.~(\ref{eom:gravitatingbrane}) 
cannot be regarded as the evolution equation for $Y_{\perp}$ 
in a usual sense. 
It will be rather appropriate to think that 
$J$ is a measure 
of the deviation from the equation of motion 
for the non-gravitating brane case. 
For the specific mode whose eigenvalue is given by Eq.~(\ref{pole:n+2}), 
we have the expression  
\begin{equation}
 J = 
   \left\{\frac{4}{\sqrt{\pi}} 
  {\Gamma\left(\nu+\frac{3}{2}\right)\over \Gamma\left(\nu\right)} 
  \left(\frac{\alb}{\ell_+}\right)^{2\nu} + O(\mu) 
                    \right\}
                {\cO}\left[\tilde\Omega\right] \,, 
\quad \mbox{for} \quad n \geq 3 \,.  
\end{equation} 
Since $\tilde\Omega$ is of $O(\mu Y_\perp)$ 
as discussed at Eq.~(\ref{Oder-relation:F-Omega-Y}), 
we can conclude that 
\begin{equation}
  J \sim O(\mu Y_\perp) \,, \quad \mbox{for} \quad n \geq 3 \,,   
\end{equation} 
for this discrete mode. 
Therefore when the weak backreaction condition $\mu \ll 1$ is satisfied, 
the source term $J$ appearing in Eq.~(\ref{eom:gravitatingbrane}) 
can be regarded as a small correction to the equation of motion  
for a non-gravitating brane~(\ref{eom:non-grav-brane}). 

In the above we have identified one discrete spectrum 
as the perturbation mode 
corresponding to the brane fluctuations. 
However, one may be still puzzled with the statement  
presented below Eq.~(\ref{eq:X-master-variable});  
Even in the limit of the weak backreaction, 
``a self-gravitating brane can oscillate only while bulk gravitational waves 
keep coming to it.'' 
We will explain that there is a loophole in this statement,  
and will find that the picture given by considering a self-gravitating brane 
is consistent with that obtained by considering a non-gravitating brane. 
First of all, even if we start with a non-gravitating brane,  
we can easily imagine that the bulk gravitational waves are emitted as a result 
of oscillation of the brane at the next level of approximation. The amplitude of the 
emitted gravitational waves will be proportional 
to the brane tension and hence proportional to $\mu$. 
As we have seen in Eq.~(\ref{Oder-relation:F-Omega-Y}), 
the amplitude of gravitational waves is smaller by factor 
$\mu$ compared with $Y_\perp$ in the description by a 
self-gravitating brane. 
Hence, there is no inconsistency as for the relative amplitudes 
between bulk gravitational waves and brane fluctuations. 
The remaining puzzle is the following. 
Although the brane described by a non-gravitating brane approximation 
can fluctuate without the incoming gravitational waves, 
it looks unable to fluctuate without incoming waves in 
the picture described by the self-gravitating brane. 
This puzzle is solved by showing 
that in fact there is a solution for the self-gravitating brane 
which satisfies no-incoming wave condition.  

To show the existence of a solution satisfying 
no-incoming wave condition, we solve the equation for the 
$\tau$-dependent part of the mode functions
(\ref{eq:T}) with $A = \cosh \tau$, 
\begin{equation}
   \left\{ \partial_\tau^2 
        + {\tilde {L}(\tilde{L}+1)\over \cosh^2 \tau} 
        + p^2 
   \right\} \left( \frac{\cT_p(\tau)}{\cosh^{\nu}\tau}\right) 
    =0 \,,  
\end{equation}
where $\tilde{L}:={L}+\nu$ with $L=0,1,2, \cdots$. 
We consider the mode function that satisfies 
the positive frequency condition,
$(\cosh \tau)^{-\nu} \:{{\cT}_p(\tau)} \sim e^{-ip\tau}$, 
at $\tau\to -\infty$. 
This mode function is given by 
\begin{eqnarray}
 (\cosh \tau)^{-\nu}{\cT}_p(\tau) & = &  
          e^{\pi p} 
          P^{ip}_{\tilde{L}}(-\tanh \tau)\cr
         & = & 
           {e^{-ip\tau} \over \Gamma(1-ip)} 
          F\left(-\tilde{L}, \tilde{L}+1, -ip+1;
                 {1\over 1+e^{-2\tau}}\right)\cr 
         & = & 
          { \Gamma(-ip) \, e^{-ip\tau} \over 
             \Gamma(-ip+\tilde{L}+1)\Gamma(-ip-\tilde{L})}
          F\left(-\tilde{L}, \tilde{L}+1, ip+1;
                 {1\over 1+e^{2\tau}}\right)\cr
&&\quad\quad + 
          { \Gamma(ip) \, e^{ip\tau} \over 
             \Gamma(\tilde{L}+1)\Gamma(-\tilde{L})}
          F\left(-ip+\tilde{L}+1, -ip-\tilde{L}, -ip+1;
                 {1\over 1+e^{2\tau}}\right) \,.
\label{cT}
\end{eqnarray}
Then, the asymptotic behavior of the mode function is given by 
\begin{eqnarray}
 (\cosh\tau)^{-\nu} {\cT}_p(\tau) & 
     \displaystyle\mathop{\sim}_{\tau\to -\infty} & 
       {e^{-ip\tau} \over \Gamma(1-ip)} 
           ,\cr
     & \displaystyle\mathop{\sim}_{\tau\to +\infty} & 
            {\Gamma(-ip)e^{-ip\tau}
       \over \Gamma(-ip+\tilde{L}+1)\Gamma(-ip-\tilde{L})} 
       + {\Gamma(ip)e^{ip\tau} \over 
         \Gamma(\tilde{L}+1)\Gamma(-\tilde{L})} \,. 
\end{eqnarray}
Choosing the time dependent part in this way, 
the solution with the eigenvalue given in Eq.~(\ref{pole:n+2}) 
asymptotically behaves as 
$\propto e^{-ip(\tau-|\zeta|)}$ at 
$\tau\to -\infty$ with $|\zeta|\sim -\tau$. 
Therefore the mode function becomes a function of $\tau-|\zeta|$  
alone. 
This means that there are no-incoming waves. 
On the other hand, at $\tau\to +\infty$ 
there are two components which behave as 
$\propto e^{-ip(\tau-|\zeta|)}$ and $\propto e^{ip(\tau+|\zeta|)}$. Hence, outgoing radiation exists. 

Finally we point out that the eigenvalue given in Eq.~(\ref{pole:n+2}) has imaginary part.  
This means that the amplitude of this solution decreases like 
$\exp[-\Im(p-i\nu)\tau]$. This decrease in amplitude can be understood 
as radiative damping effect due to gravitational wave emission.

\subsection{4-dimensional bulk ($n=2$) case} 
\label{subsection:n=2}
{}For $n=2$, there is no normalizable discrete spectrum of $\cW$ 
corresponding to the brane fluctuations. But we can still construct 
a solution which satisfies the no-incoming wave condition. 
When we choose the $\tau$-dependent part of the mode function 
as given in Eq.~(\ref{cT}), the no-incoming wave condition 
is given by $(\cosh\zeta)^{-\nu} \cQ_p=e^{ip|\zeta|}$ at
$|\zeta|\to\infty$. For $n=2$, this means that the solution 
with the no-incoming wave condition becomes 
\begin{equation} 
 \cQ_p=e^{ip\zeta}\,,\quad \mbox{for}~\zeta>\zeta_B,\quad\quad
 \cQ_p=e^{-ip\zeta} \,,\quad \mbox{for}~\zeta<\zeta_B \,.  
\end{equation}
Then from the junction condition (\ref{jc:W2}),  
we have 
\begin{equation}
      p= -i\mu \,.  
\label{pspecialn=2}
\end{equation}
For this eigenvalue, we find 
\begin{equation}
J \sim O(\mu^2 Y_\perp) \,. 
\end{equation}
As in the cases with $n\geq 3$, 
$Y_\perp$ therefore approximately satisfies the same equation as 
in the case of a non-gravitating brane. 
The only difference from the cases with $n\geq 3$ 
is that the solution is excluded from a complete set of 
solutions for $\cW$ since it is not normalizable.

\section{Green's function}   
\label{sect:green-function}

In the preceding section, we have seen that the solution 
corresponding to the brane fluctuations is not an eigenmode 
for the $n=2$ case. Thus, one may suspect whether such a 
mode is physically relevant or not. 
In this section, we examine the late time behavior of the retarded 
Green's function for the master variable ${\cal W}$. 
Recently, a similar technique was used in the context of braneworld 
cosmology~\cite{Rub,Hime}. 
We will see that 
the solution corresponding to the brane fluctuation discussed 
in the preceding section naturally dominates the late time 
behavior even if it is not contained in the eigenmodes. 

For simplicity, we concentrate on the \dsds(/) case as before. 
The Green's function for each component in the harmonic expansion  
in the $K=1$ closed chart satisfies 
\begin{equation}
\cL\, G(\tau,\chi;\tau',\chi')= -\delta(\tau-\tau')\delta(\chi-\chi') \,, 
\end{equation}
where we substitute $A = \cosh \tau$ and $k^2=L(L+n-1)$ 
in the expression of $\cL$ given in Eq.~(\ref{realmaster}). 

To find the expression for the retarded Green's function, 
we first construct $\delta(\tau-\tau')$ by a superposition of 
the product of $\tau$-dependent part of the mode functions 
given in Eq.~(\ref{cT}).  
{}From the expression in the second line of Eq.~(\ref{cT}), 
we can see that 
this function does not have a pole with respect to $p$, 
although this fact is not manifest from the expression 
in the last line.  
For the purpose of constructing $\delta(\tau-\tau')$, 
these mode functions are to be normalized with respect to the 
inner product
\begin{equation}
 ({\cal T}_{p_1},{\cal T}_{p_2}) =\int_{-\infty}^{\infty} 
  \frac{d\tau}{{(\cosh \tau)^{n-2}}}
   {\cal T}_{p_1}(\tau) {\cal T}_{p_2}(\tau) \,.  
\end{equation}
It is convenient to introduce 
$\tilde {\cal T}_p(\tau) :={\cal T}_p(-\tau)$. 
In fact, the inner product 
between ${\cal T}_p$ and $\tilde{\cal T}_{p'}$ is calculated as 
\begin{equation}
 ({\cal T}_p, \tilde{\cal T}_{p'})={2\pi i\delta(p-p')\over 
  p\,\Gamma(-ip+\tilde{L}+1)\Gamma(-ip-\tilde{L})} \,. 
\end{equation}
Then, the delta function $\delta(\tau-\tau')$ is constructed as 
\begin{eqnarray}
 \delta(\tau-\tau') &=& 
       \int_{-\infty}^{\infty} dp \: {p \over 2\pi i} \:   
       \Gamma(-ip+\tilde{L}+1)\Gamma(-ip-\tilde{L})
       \frac{{\cT}_{p}(\tau)\tilde{\cT}_{p}(\tau')}{(\cosh \tau)^{n-2}}\,. 
\end{eqnarray}
Here we have not taken into account the contribution from 
the discrete spectrum. For large negative $\tau$ or large positive $\tau'$, 
the integration contour can be closed in the upper half complex plane. 
In the upper half plane the integrand has poles at 
$p=\tilde{L}i, (\tilde{L}-1)i, \cdots, i$. 
The contribution from discrete spectrum should cancel that 
from these poles completely. 
Hence, the contribution from the 
discrete spectrum can be taken into account by shifting 
the integration contour so that it passes beyond $p=\tilde{L}i$. 

The Green's function can be constructed by assuming the following form 
\begin{equation}
 G(\tau,\chi;\tau',\chi')=\int_{-\infty}^{\infty} dp\;  
           {\cal G}_{p}(\chi,\chi') 
           {-i p \over 2\pi}
            \Gamma(-ip+\tilde{L}+1)\Gamma(-ip-\tilde{L})
         \frac{{\cT}_{p}(\tau) \tilde{\cT}_{p}(\tau')}{(\cosh \tau)^{n-2}} \,. 
\end{equation}
Then, ${\cal G}_{p}(\chi,\chi')$ should satisfy 
\begin{equation}
   \left\{ \alpha^{-(n-1)}\partial_\chi \alpha^{(n-1)}\partial_\chi 
           + \frac{1}{\al^2} \left(p^2 +\nu^2 \right) 
   \right\} {\cal G}_p(\chi,\chi')= -\delta(\chi-\chi') \,. 
\end{equation}
We denote the homogeneous solution corresponding to the 
outgoing wave at $\zeta \to \pm \infty$ by $\cQ_p^{(\pm)}(\chi)$, 
with $\zeta$ given in Eq.~(\ref{def:zeta}). Then we have 
\begin{equation}
  {\cal G}_p(\chi,\chi')={1\over W_p} 
    \left\{\cQ_p^{(+)}(\chi) \cQ_p^{(-)}(\chi') \theta(\chi-\chi')
          +\cQ_p^{(+)}(\chi') \cQ_p^{(-)}(\chi) \theta(\chi'-\chi)
    \right\} \,, 
\label{Green}
\end{equation}
with
\begin{eqnarray}
 W_p & := & 
         \{\partial_\zeta \cQ_p^{(-)}(\chi)\} \cQ_p^{(+)}(\chi) 
              -\cQ_p^{(-)}(\chi) \{\partial_\zeta \cQ_p^{(+)}(\chi)\} 
       \,.
\end{eqnarray}
The zeros of $W_p$ occurs when $\cQ_p^{(-)}$ is identical to $\cQ_p^{(+)}$. 
Namely, $W_p$ vanishes when 
the solution satisfying the outgoing boundary condition at 
$\zeta =-\infty$ simultaneously 
satisfies the outgoing boundary condition at $\zeta = \infty$.
For $n=2$, we found in Sec.~\ref{subsection:n=2} 
that the solution with the outgoing boundary condition on both sides 
satisfies the junction condition when $p=-i\mu$. 
Hence there is a zero of $W_p$ at $p=-i\mu$. 
{}For $n\geq 3$, 
when the imaginary part of $p$ is positive, 
the outgoing boundary condition implies that the wave function takes the
decaying component toward $\zeta \to \pm \infty$. Thus, 
the wave function ${\cal W}$ with the eigenvalue given in 
Eq.~(\ref{pole:n+2}) satisfies the outgoing boundary condition. 
Therefore this eigenvalue 
corresponds to a zero point of $W_p$ for $n\geq 3$.  

Let us consider the situation that a certain event 
localized in spacetime 
excites the brane fluctuations and/or the gravitational waves 
near the brane. 
This situation will be heuristically described by adding a localized source 
term on the right hand side of Eq.~(\ref{realmaster}) as 
$\cL\, \cW = \cS$. The solution is given by using the 
retarded Green's function as 
\begin{equation}
\cW=\int d^{n+2}\tilde{x}' G(\tilde{x}, \tilde{x}') \cS(\tilde{x}') \,. 
\label{formalsol}
\end{equation}  
If we consider the behavior of the perturbations at a time sufficiently late 
after the event, we will be able to close the integration contour 
in the lower half of the complex $p$-plane. 
Then, the dominant contribution 
comes from the pole of ${\cal G}_p$ that has the largest imaginary part.  
The integrand on the right hand side of Eq.~(\ref{Green}) 
has poles at $p=\tilde L i, (\tilde{L}-1)i, \cdots$ 
as well as at the zeros of $W_p$.  
The poles at $p=\tilde L i, (\tilde{L}-1)i, \cdots$ 
do not give a dominant contribution at late time 
because of the exponential suppression of $\tau$-dependent part of 
the mode function for these special values of $p$. 
As is explained above, the zero of $W_p$ with the largest imaginary part 
is nothing but the eigenfunction corresponding to the brane
fluctuations. 
Therefore the late time behavior is dominated by this eigenfunction.  
Although this eigenfunction itself has a diverging amplitude 
at $|\zeta|\to\infty$ for $n=2$, 
the solution (\ref{formalsol}) is guaranteed to go to 
zero for large $|\zeta|$ because of the causality of the retarded
Green's function. 

\section{Summary and discussion} 
\label{sect:summary}

We have studied the dynamics of perturbations of a self-gravitating vacuum 
brane of co-dimension one, taking the $(n+2)$-dimensional maximally symmetric 
configurations as the unperturbed background bulk geometry. 
In the first part of this 
paper~(Secs.~\ref{sect:background} - \ref{sect:selfperturbation}), 
we have extended the general formulas 
for the bulk and brane perturbations in terms of the gauge-invariant 
master variable, including 
the boundary conditions for the bulk perturbations 
on the brane derived from the perturbed junction conditions.  
We have derived the equation of motion for a self-gravitating brane 
in more general cases than in literature, 
confirming that the derived equation is different from that 
for a non-gravitating brane derived from the Nambu-Goto action.  
The evolution equation for the brane fluctuations 
in the case of a self-gravitating brane 
does not close by itself.  
It has a source term $J$ which is specified by the bulk metric 
perturbations. 
We have also shown that the perturbed brane motion 
$Y_\perp$ can be completely  
determined by the bulk metric perturbations. 
This fact shows that the brane does not have its own dynamical degrees of 
freedom, hence it cannot oscillate freely. 
The brane can oscillate only while incoming gravitational waves are
present. These results are generalizations 
of the works~\cite{KIF1994,II1997,II1999}, and 
they made it manifest that the way of describing dynamics of 
a self-gravitating vacuum brane looks crucially different 
from what one might naively expect from the 
analysis of a non-gravitating vacuum brane.   

In the second 
part~(Secs.~\ref{sect:fill-the-gap} and \ref{sect:green-function}), 
we have addressed a question, 
in which situation and in what manner these two ways of describing 
the brane fluctuations can be consistent, 
using the formulas developed in the first part.   
We first proposed the weak backreaction condition to clarify 
the situation in which the effect of the 
self-gravity of the brane to the bulk metric perturbations 
can be regarded as small, and hence the dynamics of a non-gravitating 
brane well approximates that of a self-gravitating one. 
The criteria for the weak backreaction condition 
we have proposed is that a non-dimensional parameter 
$\mu$ proportional to the brane tension and
the brane curvature radius, defined in Eq.~(\ref{def:mu}),  
is sufficiently small. 
This condition requires that the brane radius 
is sufficiently small compared to the bulk curvature radius 
or that the brane is an almost totally 
geodesic surface with negligible difference in the 
vacuum energies on both sides of the brane. 
 
We have examined the perturbations of the \dsds(/) system 
in the weak backreaction limit. 
We have shown 
that even when the amplitude of the incoming gravitational waves 
are of order $O(\mu)$, the amplitude of the brane perturbation $Y_\perp$ can 
be $O(1)$. 
Nevertheless a self-gravitating brane does not have 
its own dynamical degrees of freedom because the brane fluctuation 
can be entirely written by the bulk metric perturbations. 
As mentioned above, in literature we can find the argument that  
a self-gravitating brane can oscillate only while incoming 
gravitational waves are present.  

We have pointed out a loophole in this argument. If we allow the amplitude 
of perturbations to increase unboundedly in the past direction, there is 
a solution which satisfies the no-incoming wave condition. 
The solution has a discrete spectrum for $n\geq 3$, while the solution 
in the case of $n=2$ is not normalizable under the usual Hermitian measure.   
We have also shown that, when the weak backreaction 
condition holds, the gauge invariant brane fluctuations
corresponding to this solution 
approximately satisfy the same equation as for a non-gravitating brane. 
The correction due to the effect of self-gravity 
becomes small of $O(\mu Y_\perp)$ for $n\geq 3$ and 
$O(\mu^2Y_\perp)$ for $n=2$.  

Nevertheless, from the nature of the unbounded increase of  
its amplitude in the past direction, 
one might think that such a solution should be discarded 
as an unphysical one. For the case of $n=2$, the solution is 
even unnormalizable. 
However, one can see that 
this increase of the amplitude in the past direction is a 
rather expected behavior. In the treatment of the self-gravitating 
brane, we solve metric perturbations and 
brane fluctuations simultaneously. 
Therefore the leading radiation reaction is automatically 
taken into account. Hence, the radiation damping 
reduces the amplitude of the perturbation in the future direction, 
which means that the amplitude should increase toward the past direction. 

To reinforce the justification of identifying this solution 
with the one corresponding to the brane fluctuations, 
we have considered a situation in which there is an external 
force which excites the brane fluctuations. 
Using the retarded Green's function for the perturbations 
of the self-gravitating brane system, we analyzed the 
perturbations induced by an event localized in spacetime. 
We find that the late time behavior after the event 
is dominated by the solution with the no-incoming wave  
condition mentioned above. In this setup, this special solution becomes dominant 
only at a late epoch, and it is guaranteed that 
we do not see an unbounded increase of 
amplitude in the past direction because of the causality. 

In the present work, we have considered only maximally symmetric 
geometries as the background. 
It will be interesting to study less symmetric background case. 
Recently there has appeared an interesting work on the interaction between 
gravitational waves and cylindrically symmetric domain wall~\cite{N2002}, 
which also provides an example in which 
dynamics of a self-gravitating brane can be well approximated by 
a non-gravitating brane. 

In the second part of this work, we examined only \dsds(/) cases. 
An anti-de Sitter bulk spacetime plays an important role 
in the context of braneworld scenario. 
For the \adsads(/), and \minads(/) cases, however, 
the spatial section of the bulk spacetime becomes open, and 
some subtle problems might appear in the boundary condition at the infinity. 
This issue will be discussed in a separate paper. 
Studies on the co-dimension two brane and the system 
composed of many branes will be also interesting.

\section*{Acknowledgments} 
We would like to thank to J.~Garriga for useful suggestions 
and discussions. To complete this work, the discussion during and 
after the YITP workshop YITP-W-01-15 on ``Braneworld - Dynamics of 
spacetime boundary'' was useful. 
This work was supported in part by the Japan Society for the Promotion 
of Science (A.I.) and by the Monbukagakusho Grant-in-Aid 
No.~14740165(T.T.).

\section*{Appendix}  

\subsection*{A. The perturbed Einstein equations and the master equation} 

The perturbed Einstein equations on a maximally symmetric spacetime 
can be reduced to the following four gauge-invariant equations 
on the $2$-dimensional orbit space $\cN^2$ 
for modes with $k^2(k^2-nK)\ne 0$: 
\begin{eqnarray} 
&& - \square F_{ab}+D_aD_c F^c{}_b + D_b D_c F^c{}_a 
   +n \frac{D^cr}{r} \left(-D_c F_{ab} + D_a F_{cb} + D_bF_{ca} \right)  
   + \left( \frac{k^2}{r^2} -2(n-1)\lambda \right)F_{ab} 
\non \\
&& \quad 
   -D_a D_b F^c{}_c 
            - 2n \left(
                       D_aD_b + \frac{D_a r}{r} D_b+ \frac{D_b r}{r}D_a 
                 \right)F  
\non \\
&& \quad 
   -\left\{ D_cD_d F^{cd} + 2n \frac{D_c r }{r}D_d F^{cd} 
   + n(n-1)\frac{D_crD_dr}{r^2} F^{cd} 
   - 2n \square F \right. 
   - 2n(n+1) \frac{D^c r }{r}D_cF + 2(n-1)\frac{k^2-nK}{r^2}F 
\non \\ 
&& \qquad \quad 
   \left. 
   - \square F^c{}_c - n\frac{D^dr}{r}  D_d F^c{}_c 
                + \frac{k^2}{r^2}F^c{}_c - (2n+3)\lambda F^c{}_c 
   \right\} g_{ab} 
   =  0 \,,
\label{BulkPerturbationEq:scalar1} 
\\
&& 
                     - \frac{1}{r^{n-2}}D_b(r^{n-2}F^b{}_a) 
                     + rD_a \left(\frac{1}{r}F^b{}_b \right) 
                     + 2(n-1)D_a F 
  = 0 \,, 
\label{BulkPerturbationEq:scalar2} \\ 
&& 
 -\half\left\{ 
              D_a D_b + 2(n-1)\frac{D_a r}{r} D_b 
              + (n-1)(n-2)\frac{D_arD_br}{r^2}
              + 2(n-1)\frac{D_aD_br}{r} 
       \right\} F^{ab} 
\non\\ 
&& \quad 
  + \half \left\{ 
    \square  + (n-1)\frac{D^d r}{r} D_d - \frac{n-1}{n}\frac{k^2}{r^2} 
             + \lambda \right\}F^c{}_c 
   +(n-1)\left\{ \square + n\frac{D^c r }{r}D_c 
                -\frac{(n-2)}{n}\frac{k^2-nK}{r^2}
         \right\}F 
 = 0 \,,
\label{BulkPerturbationEq:scalar3}\\
&&  2(n-2)F+ F^a{}_a  = 0 \,.
\label{BulkPerturbationEq:scalar4} 
\end{eqnarray} 

{}From Eqs.~(\ref{BulkPerturbationEq:scalar2}) 
and (\ref{BulkPerturbationEq:scalar4}) we have 
\begin{eqnarray}
   D_b(r^{n-2}F^b{}_a) = 2D_a(r^{n-2}F)\,, \quad 
   F^a{}_a = -2(n-2)F \,. 
\end{eqnarray}
Then we can show that there is a master scalar $\Omega=r\tilde\Omega$ 
on $\cN^2$ by which 
$F_{ab}$ and $F$ can be expressed as 
\begin{equation}
  r^{n-2} F = \frac{1}{2n} (\Box + 2\lambda)\Omega \,, \quad 
  r^{n-2} F_{ab} = D_aD_b \Omega 
                -\left(\frac{n-1}{n} \Box + \frac{n-2}{n}\lambda \right)
                 \Omega \,,   
\end{equation} 
which are equivalent to Eqs.~(\ref{express:FabbyOmega}) 
and (\ref{express:FbyOmega}). 
Substituting these into Eq.~(\ref{BulkPerturbationEq:scalar1}), we have 
\begin{equation}
  \left( D_aD_b + \lambda g_{ab} \right) E(\Omega)=0 \,,
\end{equation}
where 
\begin{equation}
E(\Omega) := r^2\left\{ 
            \square -n\frac{D^cr}{r} D_c -\frac{k^2-nK}{r^2}-(n-2)\lambda
                \right\}\Omega \,.  
\end{equation}  
{}From this, we can finally obtain the master equation 
\begin{equation}
   \left\{
         \square - n \frac{D^cr}{r}D_c 
         - \frac{k^2-nK}{r^2} - (n-2)\lambda 
   \right\} \Omega
  =0 \,,  
\end{equation} 
which is equivalent to Eq.~(\ref{eq:reduced-master}). 
For more details of the derivation of the master equation, 
see Ref.~\cite{KIS}.

\subsection*{B. Perturbations of \dsds(3,5/4,6) system} 

Here as an exactly solvable system, we consider perturbations of 
$3,5$-dimensional de Sitter brane embedded in $4$ or $6$-dimensional 
de Sitter bulk, i.e., $n=2$ or $4$, and $\lambda_\pm = \ell_\pm^{-2}>0$.  
If we write down the radial equation by using the coordinate $\zeta$ 
defined in Eq.~(\ref{def:zeta}), we have 
\begin{equation}
  \left\{ 
         \partial_\zeta^2+p^2 + \nu(\nu-1){\alpha^2\over \ell^2} 
  \right\}
          \left(\alpha^{-\nu} \cR_p \right)=0 \,,    
\end{equation}
where $\nu=(n-2)/2$ as used in the text. 
Without loss of generality we can set the values 
of $\zeta_\pm$ on the brane to zero. 
Hence, for $n=2$ or $4$ the solution in the bulk is simply given by 
\begin{eqnarray} 
\alpha^{-\nu} \cR_p =
    \cA_\pm(p) e^{\mp ip\zeta_\pm} + \cB_\pm(p) e^{\pm ip\zeta_\pm}\,, 
\end{eqnarray}
in $\tilde \cM_\pm$. 
We quote the junction conditions (\ref{jc:2b}) and (\ref{jc:1a}) 
again as 
\begin{eqnarray} 
 \left[\partial_\zeta \cR_p\right] &=& 0\,, 
\non \\ 
 \left( p^2 + \nu^2 \right) \left[\cR_p\right] &=& 
    -2(n-1)\mu \: \partial_\zeta \cR_p\,. 
\label{app:jc}
\end{eqnarray}

\subsubsection*{\dsds(3/4) case}   

This system has been investigated in Refs.~\cite{TS1997,II1997}. 
But for the convenience for the readers we quote the results here.    
We have from the boundary conditions~(\ref{app:jc}), 
\begin{eqnarray}
\overline{\cA}&=&\overline{\cB}  \,, 
\\
\frac{ip}{2\mu} \left[ \cA + \cB \right] &=& - \cA_+ + \cB_+  \,.   
\end{eqnarray} 
Then, we obtain (see Eq.~(4.49) in Ref.~\cite{II1997}) 
\begin{equation}
   \cB_\pm = \frac{i \mu}{p + i\mu } \cA_\pm 
             +  \frac{p}{p + i\mu} \cA_\mp \,.  
\end{equation} 
We see that for the pole mode $p = -i\mu$, $\cB_\pm$ can be 
non zero even if $\cA_\pm=0$. 
Namely, there is a purely outgoing solution for $p=-i\mu$. 
For this value of $p$, the source term is given by 
\begin{equation}
   J = - \mu \frac{1}{4\alpha_B^2} 
               \left\{ 
                      5\tanh \tau {\partial \over \partial \tau} 
                      - {L(L+1) \over 2\cosh^2 \tau} 
                      + 4 \tanh^2 \tau + \mu^2 
               \right\} \tilde\Omega \,, 
\end{equation} 
where we have eliminated the second derivative of $\tau$, using 
the master equation.

\subsubsection*{\dsds(5/6) case}   

From the conditions~(\ref{app:jc}), we have 
\begin{eqnarray}
&{}&{}
  (p^2-2) \left[ \cA + \cB \right] 
  + 6ip \overline{\sqrt{1- \lambda \alpha^2} (\cA - \cB)}  
  + 3\alpha^2 \left[ \lambda(\cA + \cB)\right] =0 \,, 
\end{eqnarray}  
and 
\begin{equation}
  \left[ \sqrt{1- \lambda \alpha^2} 
     \left( \cA + \cB \right)  
  \right]
  -2ip \overline{\left( \cA - \cB \right)} 
  = 0 \,. 
\end{equation} 
{}From these, we obtain the relation between $\cA_\pm(p)$ and  $\cB_\pm(p)$ as 
\begin{equation}
   \cB_\pm = U\cA_\pm +  V \cA_\mp \,, 
\end{equation} 
with 
\begin{eqnarray}
 U &:=& {\mu \left\{ 
                       1-2p^2 +3\sqrt{1- \lambda_+ \alpha^2}
                          \sqrt{1- \lambda_- \alpha^2} 
                       -3ip \left( \sqrt{1- \lambda_+ \alpha^2}
                                  +\sqrt{1- \lambda_-\alpha^2} 
                            \right) 
                \right\}
             \over
            ip(p^2+1) - (4p^2+1)\mu - 6\mu^2ip
            -3\mu \sqrt{1- \lambda_+ \alpha^2}
                  \sqrt{1- \lambda_- \alpha^2} 
            } \,,
\\ 
 V &:=& { ip (p^2 + 1) 
             \over
            ip(p^2+1) - (4p^2+1)\mu - 6\mu^2ip
            -3\mu \sqrt{1- \lambda_+ \alpha^2}
                  \sqrt{1- \lambda_- \alpha^2} 
            } \,. 
\end{eqnarray} 
The zeros of the denominators are calculated to 
the linear order in $\mu$ as 
\begin{equation}
 p = \left( 1-\frac{3}{2} \lambda_+ \alpha^2 \mu \right) i \,, \quad 
     - \left(1 + \frac{3}{2} \lambda_+ \alpha^2 \mu \right) i \,, \quad 
     -  \left\{ 4 - 3 \lambda_+ \alpha^2 \right\}\mu i \,,  
\end{equation}
where we have used 
\begin{equation}
 \left[\sqrt{1-\lambda \alpha^2 }\right] = \left[\prtl_\chi \alpha \right] 
 = -2 \mu \,. 
\end{equation}
The eigenvalue with the 
largest imaginary part 
corresponds to that obtained in Eq.~(\ref{pole:n+2}).


\begin{thebibliography}{99}  

\bibitem{text:Polchinski} 
J.~Polchinski, 
{\em String Theory, Volume I, Volume II}, 
(Cambridge University Press, Cambridge, United Kingdom 1998). 

\bibitem{text:Vilenkin-Shellard} 
A.~Vilenkin and E.P.S.~Shellard, 
{\em Cosmic Strings and Other Topological Defects},  
(Cambridge University Press, Cambridge, United Kingdom 1994). 

\bibitem{RS1999B} 
L.~Randall and R.~Sundrum, 
{\em Phys. Rev. Lett. } {\bf 83}, 4690 (1999). 

\bibitem{VE1982}
A.~Vilenkin and A.E.~Everett, 
{\em Phys. Rev. Lett. } {\bf 48}, 1867 (1982). 

\bibitem{VEV1984}
T.~Vachaspati, A.E.~Everett, and A.~Vilenkin,  
{\em Phys. Rev. D} {\bf 30}, 2046 (1984). 


\bibitem{KIF1994}
H.~Kodama, H.~Ishihara, and Y.~Fujiwara, 
{\em Phys. Rev. D} {\bf 50}, 7292 (1994). 

\bibitem{TS1997}
T.~Tanaka and M.~Sasaki, 
{\em Prog. Theor. Phys. } {\bf 97}, 243 (1997). 

\bibitem{II1997}
A.~Ishibashi and H.~Ishihara, 
{\em Phys. Rev. D} {\bf 56}, 3446 (1997).

\bibitem{GV1991} 
J.~Garriga and A.~Vilenkin, 
{\em Phys. Rev. D} {\bf 44}, 1007 (1991).

\bibitem{Guven1993} 
J.~Guven,
{\em Phys. Rev. D} {\bf 48}, 4604 (1993).  


\bibitem{KIS} 
H.~Kodama, A.~Ishibashi, and O.~Seto, 
{\em Phys. Rev. D} {\bf 62}, 064022 (2000). 

\bibitem{Muko2000}
S.~Mukohyama, 
{\em Phys. Rev. D} {\bf 62}, 084015 (2000). 

\bibitem{II1999}
A.~Ishibashi and H.~Ishihara, 
{\em Phys. Rev. D} {\bf 60}, 124016 (1999). 

\bibitem{Rub}  
S.L.~Dubovsky, V.A.~Rubakov, P.G.~Tinyakov,
{\em Phys. Rev. D} {\bf 62}, 105011 (2000). 

\bibitem{Hime}
Y.~Himemoto, T.~Tanaka and M.~Sasaki, 
{\em Phys. Rev. D} {\bf 65}, 104020 (2002). 

\bibitem{N2002}  
K.~Nakamura, gr-qc/0205031, (2002). 


\end{thebibliography}
\end{document}